\documentclass[manuscript]{aastex}
\usepackage{emulateapj5, apjfonts, epsfig}

\def\ros{{\sl ROSAT }}

\def\chandra{{\sl Chandra }}
\def\astroh{{\sl Astro-H }}

\def\cirx1{Cir~X-1~}
\def\bx0614{4U~0614+091~}
\def\b1626{4U~1626-67~}
\def\grs1915{GRS~1915-105~}
\def\xb1822{4U~1822-37~}
\def\herx1{Her~X-1~}
\def\ergsec{\hbox{erg s$^{-1}$ }}

\def\kms{\hbox{km s$^{-1}$}}

\def\sixiii{Si~{\sc xiii}}
\def\mgxi{Mg~{\sc xi}}

\def\mgxi{Mg~{\sc xi}}

\def\sxv{S~{\sc xv}}
\def\sixiv{Si~{\sc xiv}}

\def\lapp{\ifmmode\stackrel{<}{_{\sim}}\else$\stackrel{<}{_{\sim}}$\fi}

\def\gapp{\ifmmode\stackrel{>}{_{\sim}}\else$\stackrel{>}{_{\sim}}$\fi}
\def\spose#1{\hbox to 0pt{#1\hss}}

\def\approxlt{\mathrel{\spose{\lower 3pt\hbox{$\sim$}}
        \raise 2.0pt\hbox{$<$}}}
\def\approxgt{\mathrel{\spose{\lower 3pt\hbox{$\sim$}}
        \raise 2.0pt\hbox{$>$}}}

\slugcomment{Accepted for Publication to The Astrophysical Journal}

\shorttitle{SI K EDGE STRUCTURE}

\shortauthors{SCHULZ et al.}

\begin{document}

\title{Si K Edge Structure and Variability in Galactic X-Ray Binaries }
\author{
Norbert S. Schulz,\altaffilmark{1}
Lia Corrales,\altaffilmark{1}
and
Claude R. Canizares,\altaffilmark{1}
\altaffiltext{1}{Kavli Institute for Astrophysics and Space Research, 
Massachusetts Institute of Technology,
Cambridge, MA 02139.}}

\begin{abstract}
We survey the Si K edge structure in various absorbed Galactic 
low-mass X-ray binaries (LMXBs) to study states of silicon in the 
inter- and circum-stellar medium. The bulk of these LMXBs lie toward
the Galactic bulge region and all have column densities
above $10^{22}$ cm$^{-2}$. The observations were performed with the  
\emph{Chandra} High Energy Transmission Grating Spectrometer.
The Si K edge in all sources appears at an energy value
of 1844$\pm$0.001 eV. The edge exhibits significant substructure 
which can be described by a near edge absorption feature at 1849$\pm$0.002 eV
and a far edge absorption feature at 1865$\pm$0.002 eV. Both of these
absorption features appear variable with equivalent widths 
up to several m\AA. We can describe the edge structure 
with several components: multiple edge functions, near edge 
absorption excesses from silicates in dust form, signatures from X-ray scattering optical depths,
and a variable warm absorber from ionized atomic silicon. 
The measured optical depths of the edges 
indicate much higher values than expected from atomic silicon cross sections and 
ISM abundances, and appear consistent with predictions from silicate
X-ray absorption and scattering. A comparison with models also indicates
a preference for larger dust grain sizes. In many cases we
identify \sixiii\ resonance absorption 
and determine ionization parameters between log $\xi$ = 1.8 and 2.8 
and turbulent velocities between 300 and 1000 \kms. This places the warm absorber 
in close vicinity of the X-ray binaries. In some data we observe a
weak edge at 1.840 keV, potentially from a lesser contribution
of neutral atomic silicon.
\end{abstract}


\section{Introduction\label{sec:intro}}

Our knowledge about how much dust is suspended in the gas phase of the interstellar
medium (ISM) is important for our understanding of what the ISM is composed of and
how the ISM evolves. The metal composition of the ISM helps define the cooling and star formation
rates in our Galaxy. Determining the composition of the ISM also teaches us about metal
enrichment through supernova explosions in various regions in the Galaxy. In this respect
there are fundamental differences between the Galactic disk versus the Galactic bulge.
Metal enrichment through supernova activity is thought to be very effective in the bulge 
and here we expect overabundances relative to disk compositions, specifically with respect 
to higher Z elements. This likely drives the dust-to-gas ratio in the ISM in various
regions of the Galaxy and we expect the dust depletion factors to be
different than in the Galactic disk.   

The fact that significant fractions of Mg, Si , and Fe are depleted
into dust grains in the ISM has been established in the second half of the 
last century ~\citep{draine1984, savage1996, draine2003}. 
Significant amounts of silicates are required to explain extinction
in the infrared ~\citep{draine2003}. Another issue arises about how 
silicates actually agglomerate into dust grains, which impacts
interpretations of results from X-ray dust scattering \citep{smith2008, xiang2011, seward2013})
is well as IR signatures (see \citealt{mathis1998}).  

\includegraphics[angle=0,width=8cm]{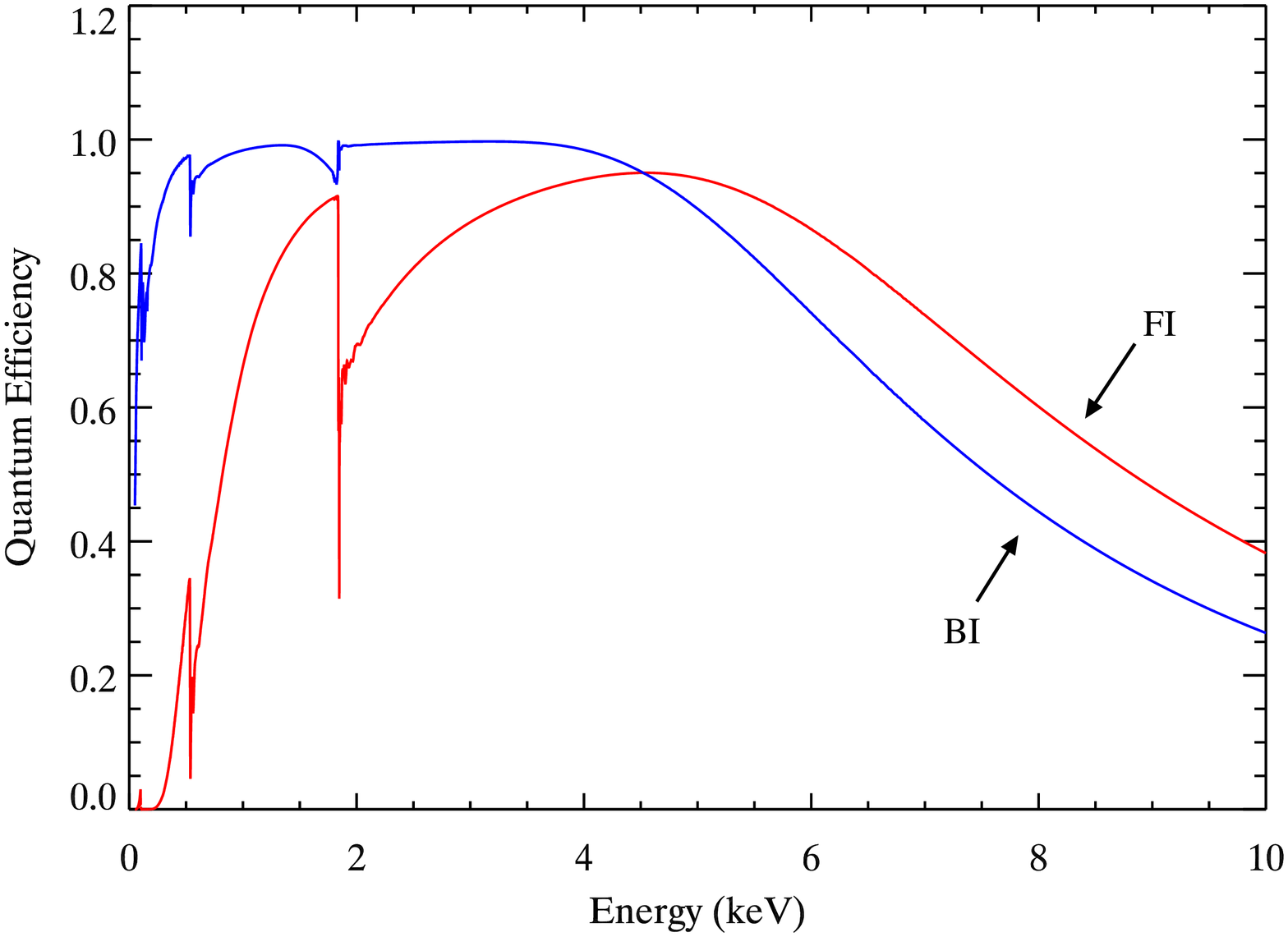}
\figcaption{The quantum efficiencies of FI (red) and  BI (blue)
devices onboard \emph{chandra}. FI devices have a substantial
instrumental Si K edge and much lower efficiences. \label{fig:qebi}}
 
Generally dust signatures in the IR to UV band are inferred indirectly through
extinction curves and to date only partially identified absorption bands.  
The X-ray band offers a rare opportunity to measure elemental dust compositions directly
through X-ray absorption fine structures (XAFS) and X-Ray absorption near edge structure (XANES).
One of the expectations with the advent of the High Energy Transmission Grating 
(HETG) spectrometer~\citep{canizares2005} was to measure precise photoelectric edges of major cosmic 
elements such as O, Ne, Mg, Si, S, Ar, Ca, and Fe. While great strides have been 
made since the launch of Chandra and XMM-Newton to determine structure and
optical depths of O K, Ne K, and Fe L edges 
~\citep{paerels2001, takei2002, schulz2002, devries2003, juett2004, juett2006}, 
studies of properties of higher Z edges have been rare to date. There 
are some individual references of silicon properties in early \chandra observations of
sources like GRS 1915+105 ~\citep{lee2002} and GX 13+1, GX 5-1, and 
GX 340+0 ~\citep{ueda2005}. The latter study specifically analyzed
X-ray binary HETG data of silicon and found a strong presence
of X-ray absorption fine structures (XAFS) in these sources. The study
concluded from these observations of GX sources that most of the 
silicon abundance exists in the form of solid silicate dust. 

\begin{figure*}[t]
\includegraphics[angle=0,width=17.5cm]{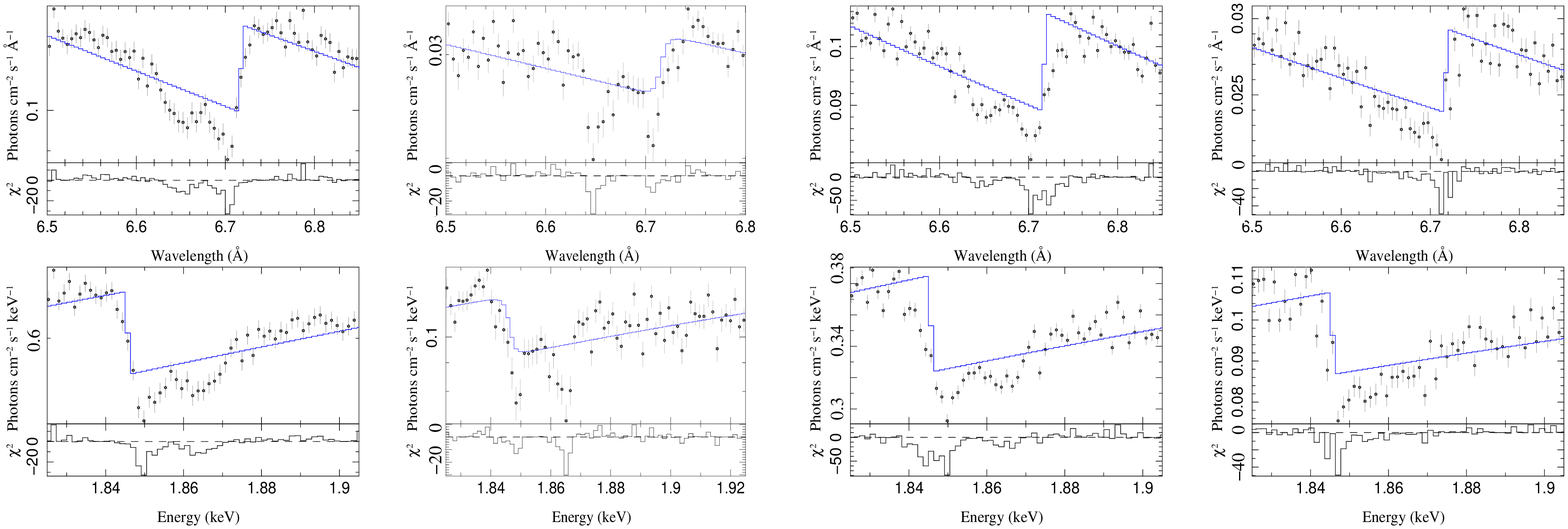}
\figcaption{Four examples of Si K edges showing edge substructure. From left to
right: GX 5-1, observation id 5888; GX 13+1, observation id 11814; GX 17+2,
observation id 6630; 4U 1728-34, observation id 6567.
\label{fig:varedges}}
\end{figure*}

In this paper we provide a comprehensive survey of the Si K edge
structure as observed the HETG spectrometer onboard \chandra. 
The survey includes 45 observations of 11 sources with 
column densities larger than
about $10^{22}$ cm$^{-2}$ and X-ray fluxes between 40 and 1100 mCrab.
The study is limited to the Si K edge region and the determination
of the edge energy and optical depth as well as any accompanying
absorption components. We specifically focus on edge location, optical
depths, common discrete absorption features, and variability.

\section{Chandra Observations\label{sec:observ}}

To date there are 45 HETG observations in the ~\chandra archive of sources which
have column densities near or above $10^{22}$ cm$^{-2}$ and sufficient
X-ray flux to study the Si K edge region. These observation are
summarized in Table 1. The sample involves 11 Galactic
sources with several observations for each source ranging
from one observation for 4U 1624-49 up to eight observations for GX 349+2 
in various data modes. 

All observations were processed using CIAO4.6 with the most recent
CALDB products and the on-line transmission grating catalog
(TGCAT\footnote{see \url{http://tgcat.mit.edu/}} 
~\citet{huenemoerder2011}) processing procedures. The zero-order point spread function (psf)
is heavily piled up in most observations and an improved zero-order position was determined
using \emph{findzo.sl} which uses the intersection of the psf
read-out streak and the HETG dispersion tracks
\footnote{see also \url{http://asc.harvard.edu/ciao/threads/}}.

\begin{table*}
\vbox{ 
\small 
\begin{center} 
{\sc TABLE~1: OBSERVATIONS AND SOURCE PARAMETERS:\label{tab:observations}} 
\vskip 4pt 
\begin{tabular}{llcccccccccc} 
\hline 
\hline 
 SOURCE &  OBSID & Date & Start & Exp. & Rate&Mode&Ftime & RA & DEC & NH & F$_x$\\ 
        &        & UT   &  UT   & ks   &  c/s&     & s & deg & deg & (1) & (2) \\ 
\hline 
 & & & & & & & & & & & \\ 
GX 5-1 & 716 & 2000-07-18 & 06:32:12 & 8.9 & 67.1 & TE & 1.328 & 270.284 & -25.079 & 5.78 $\pm$ 0.65 & 0.43  \\ 
  & 5888 & 2005-10-30 & 02:03:55 & 45.0 & 65.1 & CC & 0.003 & & & 5.76 $\pm$ 0,68 & 3.00  \\ 
  & 10691 & 2009-02-10 & 05:30:21 & 8.0 & 61.8 & CC & 0.003 & & & 5.67 $\pm$ 0.95 & 2.92  \\ 
  & 10692 & 2009-05-14 & 09:10:17 & 8.1 & 68.9 & CC & 0.003 & & & 5.36 $\pm$ 0.91 & 3.05  \\ 
  & 10693 & 2009-05-24 & 21:29:26 & 8.1 & 51.4 & CC & 0.003 & & & 6.05 $\pm$ 1.11 & 2.81  \\ 
  & 10694 & 2009-10-27 & 22:25:49 & 5.0 & 50.2 & CC & 0.003 & & & 5.61 $\pm$ 1.14 & 2.75  \\ 
GX 13+1 & 2708 & 2002-10-08 & 11:12:56 & 29.3 & 13.8 & TE & 1.094 & 273.632 & -17.157 & 4.30 $\pm$ 0.51 & 0.74  \\ 
  & 11814 & 2010-08-01 & 00:31:31 & 28.1 & 11.7 & TE & 1.094 & & & 4.54 $\pm$ 0.60 & 0.62  \\ 
  & 11815 & 2010-07-24 & 05:46:27 & 28.1 & 14.6 & TE & 1.094 & & & 4.63 $\pm$ 0.53 & 0.75  \\ 
  & 11816 & 2010-07-30 & 14:47:25 & 28.1 & 14.2 & TE & 1.094 & & & 4.65 $\pm$ 0.53 & 0.74  \\ 
  & 11817 & 2010-08-03 & 10:12:10 & 28.1 & 13.5 & TE & 1.093 & & & 4.64 $\pm$ 0.51 & 0.62  \\ 
  & 11818 & 2010-08-05 & 14:09:39 & 23.0 & 17.2 & CC & 0.003 & & & 4.22 $\pm$ 1.07 & 0.76  \\ 
  & 13197 & 2011-02-17 & 17:57:04 & 10.1 & 17.1 & CC & 0.003 & & & 4.32 $\pm$ 1.66 & 0.71  \\ 
GX 9+1 & 717 & 2000-07-18 & 03:20:24 & 8.4 & 23.0 & TE & 1.647 & 270.385 & -20.529 & 1.72 $\pm$ 0.30 & 1.00  \\ 
  & 6569 & 2007-05-11 & 19:41:36 & 33.6 & 36.8 & CC & 0.003 & & & 2.02 $\pm$ 0.28 & 1.12  \\ 
GX 349+2 & 715 & 2000-03-27 & 18:25:04 & 9.4 & 40.8 & TE & 0.9375 & 256.435 & -36.423 & 2.02 $\pm$ 0.90 & 1.57  \\ 
  & 3354 & 2002-04-09 & 00:43:26 & 25.7 & 25.8 & TE & 1.375 & & & 1.98 $\pm$ 0.65 & 1.58  \\ 
  & 6628 & 2006-07-04 & 18:52:16 & 12.5 & 77.0 & CC & 0.003 & & & 2.52 $\pm$ 0.29 & 2.33  \\ 
  & 7336 & 2006-08-20 & 05:36:32 & 12.1 & 68.2 & CC & 0.003 & & & 2.52 $\pm$ 0.28 & 1.41  \\ 
  & 12199 & 2011-10-09 & 04:31:35 & 19.2 & 42.2 & CC & 0.003 & & & 2.44 $\pm$ 0.29 & 1.92  \\ 
  & 13220 & 2011-07-05 & 20:04:36 & 18.9 & 43.6 & TE & 0.9375 & & & 2.18 $\pm$ 0.25 & 0.88  \\ 
  & 13221 & 2011-05-15 & 04:43:48 & 23.0 & 46.6 & CC & 0.003 & & & 2.35 $\pm$ 0.27 & 1.50  \\ 
  & 13222 & 2011-10-18 & 08:42:43 & 32.5 & 48.7 & CC & 0.003 & & & 2.44 $\pm$ 0.26 & 1.52  \\ 
GX 17+2 & 4564 & 2004-07-01 & 11:35:16 & 30.1 & 41.2 & CC & 0.003 & 274.006 & -14.036 & 3.64 $\pm$ 0.38 & 1.64  \\ 
  & 6629 & 2006-05-10 & 01:59:44 & 23.6 & 43.5 & CC & 0.003 & & & 3.55 $\pm$ 0.37 & 1.76  \\ 
  & 6630 & 2006-08-19 & 22:26:31 & 24.0 & 40.9 & CC & 0.003 & & & 3.58 $\pm$ 0.39 & 1.54  \\ 
  & 11088 & 2010-07-25 & 08:18:44 & 29.1 & 29.5 & TE & 1.125 & & & 3.75 $\pm$ 0.38 & 1.27  \\ 
  & 11888 & 2010-07-24 & 14:10:03 & 4.6 & 32.2 & CC & 0.003 & & & 4.03 $\pm$ 0.50 & 1.61  \\ 
GX 340+0 & 1921 & 2001-08-09 & 05:36:19 & 23.4 & 8.4 & TE & 1.25 & 251.449 & -45.611 & 6.31 $\pm$ 1.05 & 0.64  \\ 
  & 1922 & 2001-08-09 & 12:35:11 & 5.8 & 12.8 & CC & 0.003 & & & 6.81 $\pm$ 0.99 & 0.90  \\ 
  & 6631 & 2006-05-19 & 20:11:58 & 25.0 & 20.7 & CC & 0.003 & & & 6.84 $\pm$ 0.78 & 0.13  \\ 
  & 6632 & 2006-06-21 & 10:08:31 & 23.6 & 19.9 & CC & 0.003 & & & 6.61 $\pm$ 0.77 & 0.13  \\ 
4U 1705-44 & 1923 & 2001-07-01 & 17:20:28 & 24.4 & 18.5 & TE & 1.25 & 257.227 & -44.102 & 3.11 $\pm$ 0.35 & 0.79  \\ 
  & 1924 & 2001-06-25 & 00:38:41 & 5.9 & 25.9 & CC & 0.003 & & & 3.46 $\pm$ 0.60 & 0.91  \\ 
  & 5500 & 2005-10-26 & 07:20:13 & 26.5 & 5.8 & TE & 1.25 & & & 3.86 $\pm$ 0.39 & 0.17  \\ 
4U 1728-34 & 2748 & 2002-03-04 & 15:19:31 & 29.6 & 3.4 & TE & 1.25 & 262.990 & -33.834 & 5.54 $\pm$ 0.62 & 0.17  \\ 
  & 6567 & 2006-07-18 & 19:51:21 & 151.25 & 11.1 & CC & 0.003 & & & 4.41 $\pm$ 0.47 & 0.42  \\ 
  & 6568 & 2006-07-17 & 12:42:01 & 49.3 & 9.0 & CC & 0.003 & & & 4.44 $\pm$ 0.48 & 0.33  \\ 
  & 7371 & 2006-07-22 & 21:46:23 & 39.6 & 6.3 & CC & 0.003 & & & 4.45 $\pm$ 0.54 & 0.23  \\ 
4U 1624-49 & 4559 & 2004-06-04 & 06:25:02 & 73.4 & 1.4 & TE & 1.6 & 247.010 & -49.190 & 6.47 $\pm$ 0.72 & 0.11  \\ 
GX 3+1 & 2745 & 2002-04-09 & 10:50:50 & 8.9 & 29.6 & CC & 0.003 & 266.983 & -26.564 & 3.40 $\pm$ 0.50 & 0.99  \\ 
  & 16307 & 2014-07-22 & 23:36:59 & 43.6 & 18.8 & TE & 1.094 & & & 3.47 $\pm$ 0.38 & 0.71  \\ 
  & 16492 & 2014-08-17 & 00:05:52 & 43.6 & 19.2 & TE & 1.094 & & & 3.55 $\pm$ 0.38 & 0.71  \\ 
Ser X-1 & 17485 & 2015-02-13 & 08:08:40 & 83.2 & 22.3 & TE & 0.42 & 279.990 & +05.036 & 1.23 $\pm$ 0.13 & 0.54  \\ 
  & 17600 & 2015-02-21 & 06:36:50 & 37.1 & 21.7 & TE & 0.42 & & & 1.42 $\pm$ 0.17 & 0.46  \\ 
& & & & & & & & & & &   \\ 
\hline 
\end{tabular} 
\end{center} 
(1) 10$^{22}$ cm$^{-2}$;  (2)  10$^{-8}$ erg s$^{-1}$ sec$^{-2}$ 
\normalsize 
} 

\end{table*}

The spectral analysis was performed using the latest version of
\emph{ISIS}\footnote{see \url{http://space.mit.edu/ASC/ISIS}} with
imported \emph{Xspec.v12} functions for spectral
modeling. Uncertainties are 90$\%$ confidence limits calculated using
the multi-parameter grid search utility \emph{conf$\_$loop} in
\emph{ISIS} ~\citep{houck2000}.
In all cases we had sufficient statistics to
pursue $\chi^2$ minimization. In the case of searching for more detailed
edge structure we also employed a cash statistic in combination with
a subplex fit method as provided by \emph{ISIS}.

\subsection{Data Modes and Reduction 
\label{sec:data}}

Table 1 shows that about 40$\%$ of the observations were taken in timed event
(TE) mode. Some of these data were taken in the early phases of the 
mission where the effects of pileup in the HETG has not yet fully been 
recognized and mitigated.
Specifically in observations IDs $<$ 1000, photon pileup in grating spectra is significant and
even though there are procedures to reconstruct the data, these results 
have high systematic uncertainties. We include these data in the survey but
treat them with caution. Most TE mode data were taken with tight subarrays
with row numbers between 300-420 and consequently much lower frame times. 
The bulk of the pileup events comes from
the MEG +1st order where the Si K edge resides on a back-illuminated CCD. These devices have 
significantly higher effective areas between 1 and 3 keV than front-illuminated devices and 
in TE mode we exclude some of these data (see Fig.~\ref{fig:qebi}). We can in extreme cases eliminate
significant sources of pileup by only using HEG +/-1st orders. 
The optical depths in the TE mode sample are then affected up to $10\%$ by pileup at the Si K edge
in the worst cases.

In most cases, however, pileup affects the edge by much less than $10\%$. We tested this
on the example of GX 13+1 using the \emph{gpile2} function in \emph{ISIS}, which
allows a simple pileup reconstruction of HETG continuum spectra. Even though here we
find pileup of up to 20$\%$ in the hard parts of the spectra $< 4$~ \AA, the value around
Si K was of the order of $8 \%$. The continua in LMXB spectra increase with energy in part
because of absorption due to high column densities but also intrinsically. 
Thus even though the HETG 
effective area is high around Si, the bulk of the photon pileup affects the hard part
of the spectra. Furthermore, an edge optical depth is only affected by the pileup difference
between the top and the bottom edge flux. In GX 13+1 the difference of the effective pileup
at the top and bottom of the Si K edge is less than $5 \%$. The average effect in all sources
is $5 \%$ and we correct optical depths upwards by this amount as well as add this amount to the 
uncertainties in optical depth and equivalent width. 

The remainder of the data were taken in continuous clocking (CC-mode) 
with a fast read time (note: the frame times 
in Table~1 are approximate and for reference only). Here the spectra are devoid of 
any pileup and ideally the edge structure and optical depth are fully preserved.
However, absorbed CC-mode spectra are affected by contributions from dispersions
from all incident photons. In most cases this at least includes the central source but
also a substantial scattering halo. In CC-mode these contributions collapse into 
a one dimensional image and the HETG extraction then contains the source spectrum as well
as dispersed photons from the zero order scattering halo which get added to the 
dispersed first order spectra and ultimately will alter the spectral shape of these spectra. 
In addition, while in TE- and CC-mode
any contributions including the ones from higher orders are effectively filtered out by the 
pulse height channels in every CCD pixel, this filter is not fully effective in CC-mode. 
For example, the HEG 1st orders are confused by the MEG 2nd orders. 
In the sources of this sample the scattering halo is prominent because of the 
high absorption columns involved. While the edge structure remains preserved,
optical depth values become affected. From comparisons with data taken in
TE-mode we estimate that this effect adds  
about 20$\%$ systematic uncertainty to the optical depth on average. 

FI devices also have an intrinsic Si K edge from the SiO$_2$ gate structural layer located
at the atomic Si K edge location. A high resolution correction is applied in TE-mode
data to account for this contribution. However, since the optical depths appear 
slightly reduced in CC-mode due to the effects described above, this correction cannot
be properly applied and we cannot use FI data. 
In BI devices there is no significant instrumental
contribution to the Si K edge (see Fig.~\ref{fig:qebi})and in CC-mode we therefore use exclusively the 
MEG +1st order located on the S3 device. 
We then correct optical depths by 20$\%$ as well as add this amount to the
uncertainties in optical depth and equivalent width.

\subsection{Survey Source Parameters
\label{sec:survey}}
The spectra
were all fitted with a two component spectral model consisting of a 
disk blackbody in the soft and a power law in the hard end of the 
spectral bandpass plus absorbing column density for which we used the 
\emph{TBnew} function in \emph{XSPECv12}. For this continuum
fit we apply the ISM abundances from ~\citet{wilms2000}. 
The bandpass is limited at the soft end by the high 
column density to about 9\AA\ (1.4 keV) and at the high end to 5.2~\AA\ (2.6 keV) in CC-mode 
spectra leaving about 180 independent resolution bins for the spectral fitting
in MEG spectra, and 340 bins in HEG spectra.  
The spectra are well fit by this model with reduced $\chi^{2}$ values below 1.8 in the 
worst case. Note, in this analysis we do not require a physical interpretation
of the spectral continuum, but want a good representation for flux and broadband
column densities. The resulting hydrogen equivalent column densities are listed in Tab.~1. 
We will use this column as the x-scale in some of the plots.
All sources
are bright with 1st order count rates between as low as 1.4 cts/s to as high as 77 cts/sec, 
which translates to a flux range of 1$\times 10^{-9}$ \ergsec to 3 $\times 10^{-8}$ \ergsec  
(0.04 to 1.1 Crab) in the range between 2 and 10 keV.
The column densities in the data vary from
1.23$\pm0.13 \times 10^{22}$ cm$^{-2}$ in obsid 17485 for Ser X-1 and 
6.84$\pm0.92 \times 10^{22}$ cm$^{-2}$ in obsid 6631 for GX 340+0. 
The columns have low statistical but a substantial systematic
error of 10$\%$ added due to the limited bandpass. Most values obtained are consistent
within these uncertainties with columns obtained in previous studies.

\includegraphics[angle=0,width=8cm]{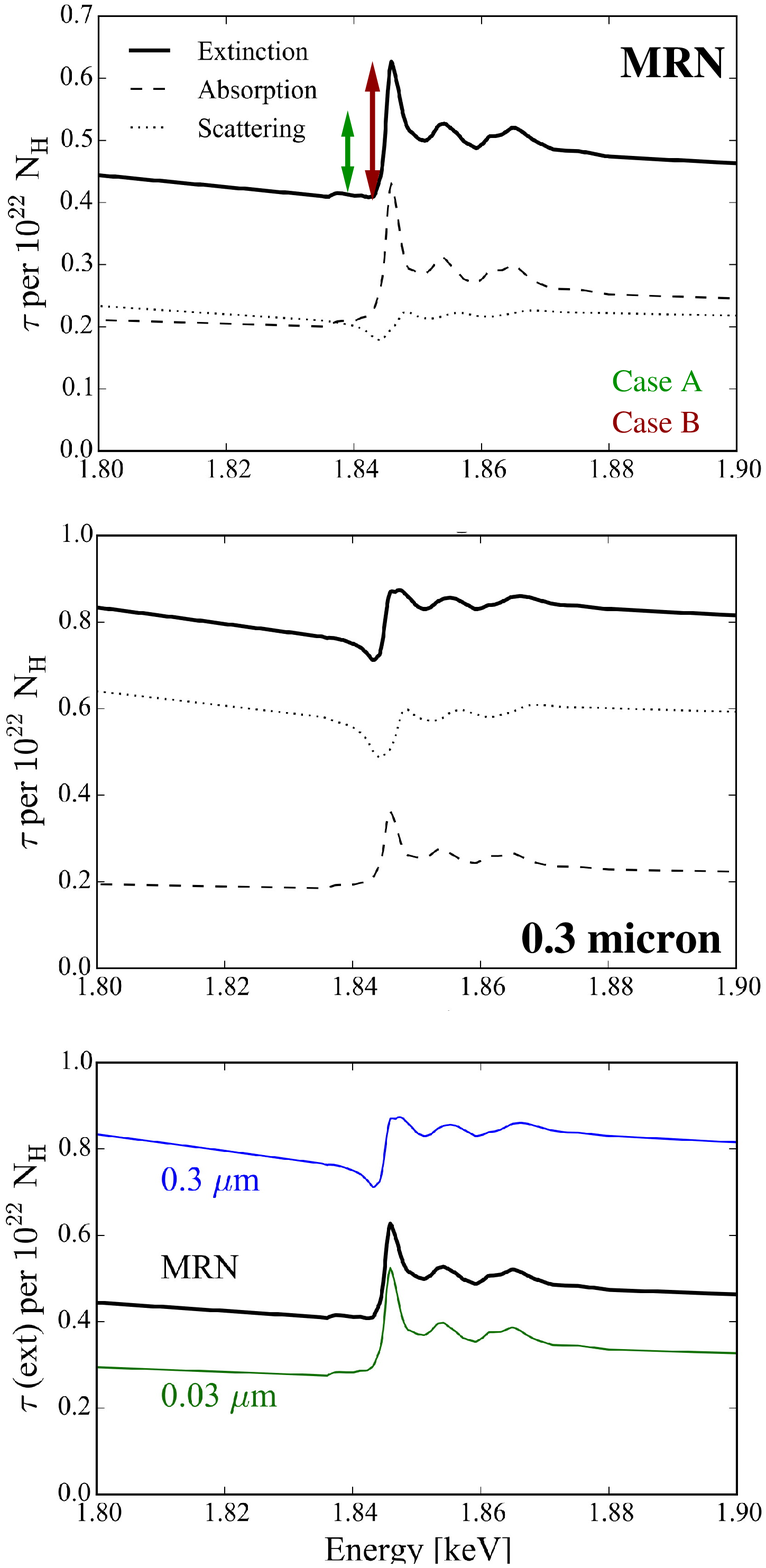}
\figcaption{The Si K optical depths from the ~\citet{draine2003} model calculated for
the MRN grain size distribution (top) and for large grain sizes
of 0,3 $\mu$m (middle) from dust codes in ~\citet{corrales2015}.
Shown are absorption and scattering  as well as the final extinction functions.
The red and green double arrows mark the optical depth extractions for cases A and B. A comparison
of various grain sizes (bottom) shows that X-ray extinction at Si is very similar for the MRN
distribution and small grain sizes, while larger sizes behave significantly different
due to increased contributions from scattering.
\label{fig:lia_sik}}

\subsection{Survey Analysis Strategy
\label{sec:strategy}}

The variety of data modes and their systematic uncertainties do not allow to 
directly fit models to these data to determine sensible physical information. However,
the data are good enough to deduce common trends with respect to existing models. First, 
we take advantage of the superbly calibrated energy (wavelength) scale of the 
\emph{Chandra} HETG data. The scale calibration is entirely independent of the 
data modes and have low systematical uncertainties for all survey data. In this 
respect we will determine most accurate energy (wavelength) locations of Si K edge 
features with mosty statistical variances of the order of about 1 eV or 4 m\AA. 

Second, the optical depths of features in the Si K edge is a directly measurable quantity
fairly independent of the detailed shape of the edge but still subject to systematical
uncertainties that come with the use of various data modes. However in  the case of
the optical depths we can account for most of them. In this paper we therefore 
survey the data in terms of how optical depths of Si K absorption features distribute
among X-ray sources in the Galactic bulge and how these properties compare to
predictions from existing models of atomic and silicate absorption models.

\section{Edge Models}

There are several ways to describe neutral absorption edges of elements and molecular
compounds. These range from the standard edge model usually describing elements in their
basic atomic form up to more complex structure: low ionization levels as
shown  for O-K  and Ne-K ~\citep{juett2004, juett2006}, multiple edges from unresolved
resonances as in Fe-L ~\citep{kortright2000, schulz2002}, and complex structures 
from laboratory measurements as applied in ~\citet{ueda2005} for the cases of magnesium,
silicon, and sulfur K edges. 
In the following we describe the models we use for our Si K data in detail.
 
\subsection{Standard Edge Absorption}
\label{sec:standard}

In order to explore a first cut of the structure of the Si K edge, we employ the standard edge
function implemented in \emph{XSPECv12}.
The K edge structure can be decomposed into various elements. Standard edge parameters
are the edge location ($\lambda_{SiK}$ [\AA], $E_{SiK}$ [keV]) and optical depth $\tau$ 
giving it a step like shape. The function in detail is

\begin{equation}
F(E)  = 1~~~~~~  E < E_{SiK} , ~~~F(\lambda) = 1~~~~~~  \lambda > \lambda_{SiK}  \nonumber
\end{equation}
\vspace{-4mm}
\begin{eqnarray}
F(E)       & = & exp(-\tau (E/E_{SiK})^{-3})~~~~~~~   E > E_{SiK}  \nonumber 
\end{eqnarray}
\vspace{-4mm}
\begin{eqnarray}
F(\lambda) & = & exp(-\tau (\lambda_{SiK}/ \lambda)^{-3})~~~~~~~  \lambda < \lambda_{SiK} \nonumber 
\end{eqnarray}
 
\noindent
with an exponential edge recovery at the high energy side.
The edges in the survey are also accompanied by structure, which in all cases
are represented as Gaussian shaped absorption features. Figure~\ref{fig:varedges}
shows edge structure examples in four different sources plotted on an ~\AA\ (top)
and keV (bottom) scale. The following subsections
describe these structures in detail for all survey sources.

\subsection{Extinction from Silicate Dust}
\label{sec:silicate}

Most previous efforts to decribe neutral and near neutral absorption edges in the 
X-ray band relied entirely on atomic absorption properties at or in the vicinity of the edge.
In the case of silicon this approach is incomplete since it is now assumed that 
at least a significant fraction of Si is bound in molecular form and conglomerated
into dust grains. In that case X-rays in the line of sight are affected by scattering as
well as absorption, both contributing to the affecting optical depth. The sum of the 
absorption and scattering optical depth is called extinction and affects radiation
throughout the entire electromagnectic spectrum. Studies by ~\citet{weingartner2001}, 
~\citet{draine2003} and most recently ~\citet{hoffman2015} and ~\citet{corrales2016}, 
laid out some of the 
ground work how to treat the Si K edge band with respect to both scattering and 
absorption. 

Figure~\ref{fig:lia_sik} shows the optical depth from the \citet{draine2003} model per unit column for
the Si K edge for scattering, absorption and total extinction
(see also ~\citealt{corrales2016}). Using the dust scattering codes of ~\citet{corrales2015}, the
total Si K edge extinction cross section shown in Figure~\ref{fig:lia_sik} (top) was calculated
from a power law distribution (MRN) of dust grain sizes (see ~\citealt{mathis1998}), 
$dn/da \propto a^{-3.5}$ where $a$ is the
radius for spherical dust grains. The minimum and maximum cut-offs to the grain size
distribution were $0.005~\mu{\rm m}$ and $0.25~\mu{\rm m}$, respectively. To calculate
the Mie scattering cross section, we use the optical constants derived by \citep{draine2003},
which rely on laboratory Si K edge absorption measurements of silicate materials~\citep{li1995},
in this case specifically modeling properties of olivines, i.e. Mg$_2$SiO$_4$.
Figure~\ref{fig:lia_sik} (middle) shows the same but for 0.3 $\mu$m grain sizes only.
To arrive at the total opacity of the Si K edge, Figure~\ref{fig:lia_sik} assumes that
all neutral Si is locked up in dust and that
all of the interstellar dust is in amorphous silicate (olivine) form. These calculations
also were done using solar abundances~\citep{asplund2009} and we use a renormalization
factor (0.59 for olivine) to convert to ISM abundances~\citep{wilms2000} used in the data analysis.

\vspace{0.2cm}
\includegraphics[angle=0,width=8cm]{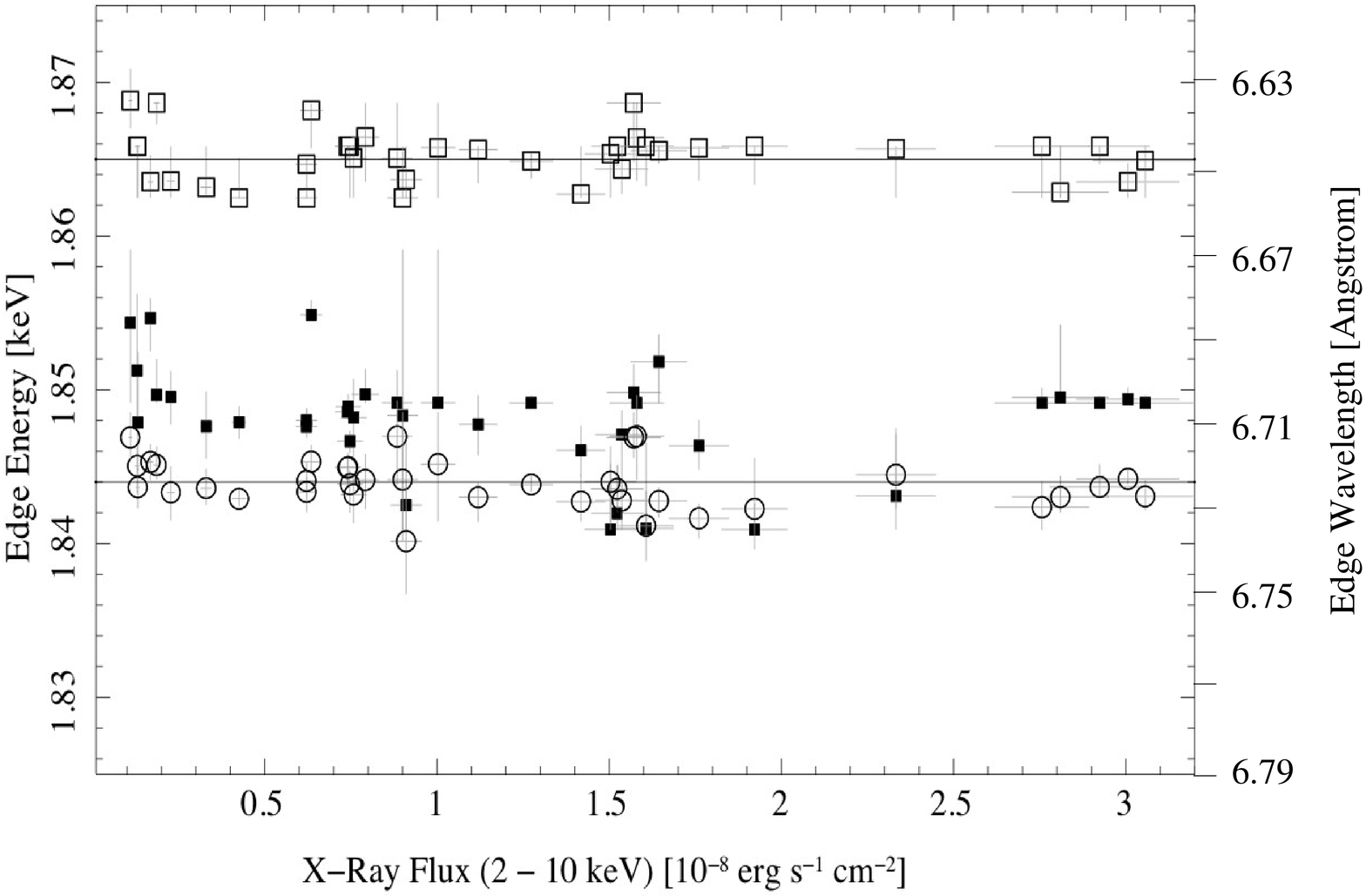}
\figcaption{The measured Si K edge locations of all survey sources as determined
by the standard edge model (unfilled circles). The lower straight line marks
the average value at 1.844$\pm0,001$ keV. Also plotted are the centroid values
of the near (filled squares) and far edge (unfilled squares) absorption features
where they were detected. The upper straight line marks the rest frame
value of the \sixiii\ resonance line. A wavelength scale is added on the right side.
\label{fig:standard}}
\vspace{0.2cm}

The fit with the standard edge function determines the optical depth of the
Si K edges in all survey observations. This optical depth ignores additional edge structure
and treats the Si K edge as the simple step function determined by the continuum
below and above the edge location (in keV units). We refer to these depths as \emph{case A}. Another
method to determine the optical depth is to refer to the flux values directly below and above
the edge location (in keV units), which accounts for the substructure - in this case the near edge absorption
(see Sect.~\ref{sec:nearedge}) - of the edge. Here the optical depth is calculated
by

\begin{equation}
\tau = ln (\frac{f_a}{f_b}),
\end{equation}

\noindent
where $f_b$ is the flux below, $f_a$ the flux above the edge location where
it is lowest (with respect to energy scale).
We refer to these depths as \emph{case B}. 

\section{Measuring the Si K Edge Structure\label{sec:edge}}

\subsection{Edge Location 
\label{sec:value}}

\begin{figure*}
\includegraphics[angle=0,width=18cm]{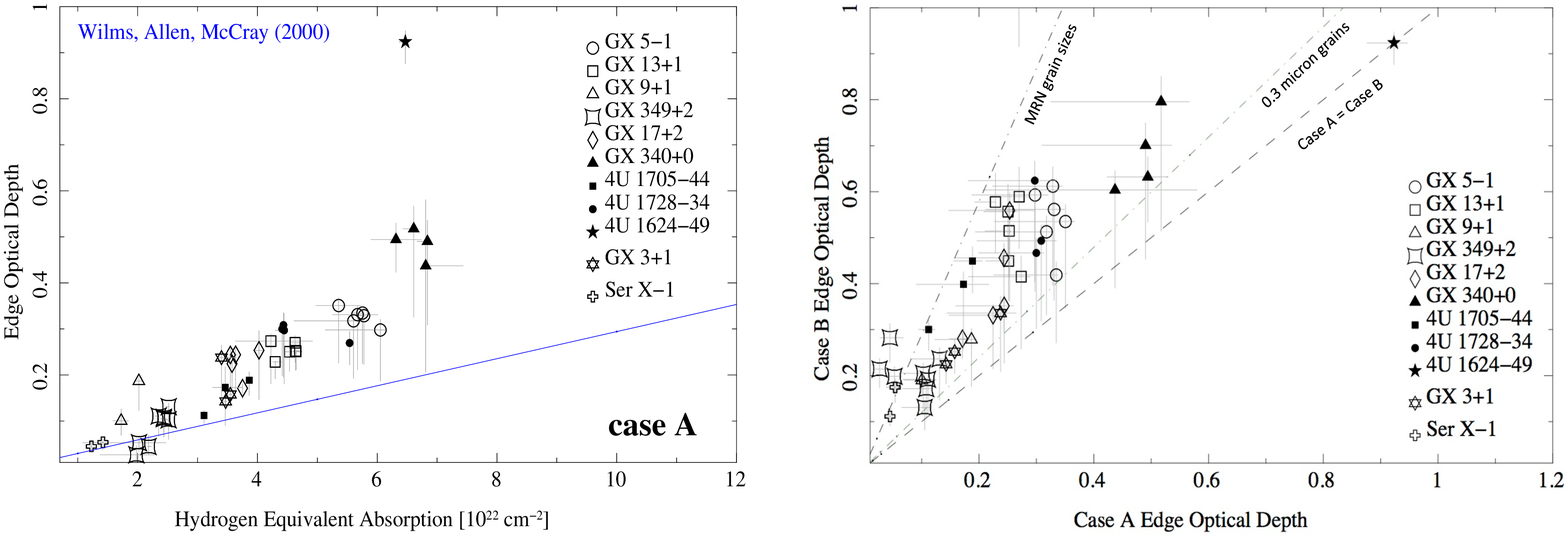}
\figcaption{{\bf Left:} The Si K edge optical depth for \emph{case A} plotted against the broadband determination
of the column density for all survey observations. The blue line represents the
expectation from ~\citep{wilms2000}. {\bf Right:} \emph{Case A} optical depths plotted
against \emph{Case B} optcial depths as defined in Fig.~\ref{fig:lia_sik}. Marked are
model cases: the standard edge model where case A and case B are the same,
and two grain models for MRN grain sizes and 0.3 $\mu$m grain sizes.
\label{fig:depth}}
\end{figure*}

We fitted all Si K edge data with the standard edge model described in Eq. 1. 
The unfilled circles in  Figure~\ref{fig:standard} show the measured Si K edge 
energies $E_{SiK}$ for
all survey sources as a function of source flux. The measured edge energies show
little variations, the smallest value we find is at 1.840 keV, the largest at 1.848 keV, both these
values come from edges with lower statistics. Edge values cluster tightly around an average value of
$E_{SiK} = 1.844\pm$0.001 keV ($\lambda_{SiK} =6.724\pm$0.004~\AA). 
Many of these edges have additional structure
and the edge fit here does not take in account any contribution from the near
edge absorption feature (Sect.~\ref{sec:nearedge}) which adds a systematic
uncertainty of another 0.75 eV (0.003~\AA). Variations within single sources 
are very small, the largest deviations are again associated with lower statistics. 
We do not
observe any dependence of the edge location with source flux, nor any correlations with
other source or edge parameters. The expected atomic edge location is 1.839 keV~\citep{bearden1967} and 
the measured values are larger by 5 eV, which is an order of magnitude larger than
the uncertainty of the HETG energy scale.

\subsection{Optical Depth                   
\label{sec:optical}}

The hydrogen equivalent column density $N_H$ for a specific element $z$ relates 
to the optical depths $\tau_z$ via the 
absorption cross sections $\sigma_z$ as

\begin{equation}
\tau_z = N_H \times A_z \times a_z \times \sigma_z,
\end{equation}

\noindent
where $A_z$ the abundance from a specified
distribution and $a_z$ is an over-abundance factor 
Here we generally refer to ISM abundances from ~\citet{wilms2000}.
The atomic absorption cross section $\sigma_z$ in that study has a value
of 1.58$\times10^{-19}$ cm$^2$ from \citet{verner1995}. 

Case A depths (see definition in Fig.~\ref{fig:lia_sik}) show that the measured optical depths in most sources
are significantly higher than expected from the ISM abundance and the atomic Si cross section
(blue line), on average 
about a factor 2. For all sources we obtain $a_z$ values $\sim 2$ for Si abundance
factors when we fit the edges with the \emph{TBnew} function and a free Si abundance parameter.
We will show below that that much of this apparent overabundance is removed when we use better Si 
cross sections. 
The scatter of the data within each of the sources (i.e. same symbols) is mostly due to the 
different data modes. 
Case B depths (see Fig.~\ref{fig:lia_sik}) of all sources 
are much larger by factors 1.6 (GX 349+2) up to a factor 5 (GX 13+1) (Figure~\ref{fig:depth}).
The plot on the left shows the case A optical depths versus the fitted broadband
column densities. Even though there is  some scatter in the data, the optical depths
show a good correlation with overall column. The plot on the right shows the 
case A optical depths versus the case B optical depths.
The scatter of case B optical depths  
is somewhat larger. This is likely due to systematic variations
from the different data modes (see Sect. 2.1) as well as 
variability in the 
near edge absorption features (see below). 

These depth excesses cannot be attributed to any of our measuring deficiencies as 
either pileup in TE mode as well as imaging distortions in CC mode would reduce our
edge optical depth determinations. 
Note, 4U 1624-49 is exceptional in that the depth is higher by about a factor
5 for case A, but here the case for substantial intrinsic 
absorption is known~\citep{xiang2009}. 
Here it is also interesting to note that the data do not show any near
edge absorption (case A = case B).

\subsection{Near Edge Absorption\label{sec:nearedge}}

The Si K edge structures in Figure~\ref{fig:varedges} show significant
absorption close to the bottom of the edge function. We modeled the properties
of this absorption by fitting a broadened Gaussian function to the data. 
The filled squares in Figure ~\ref{fig:standard} show the centroids of 
the absorption line fits. Their
values lie between 1.844 and 1.852 keV with the bulk fit at 1.849+\-0.002 KeV
(6.706$\pm$0.006~\AA). In some cases this absorption feature has very low contrast with respect to 
the edge function and the centroid fit overlaps with the actual edge location. In the MEG, where we
have most of our data, we observe this as a smeared but steep and unresolved edge drop. The average
offset of the absorption centroid with respect to the edge location is about 5 eV.

\begin{figure*}
\includegraphics[angle=0,width=18cm]{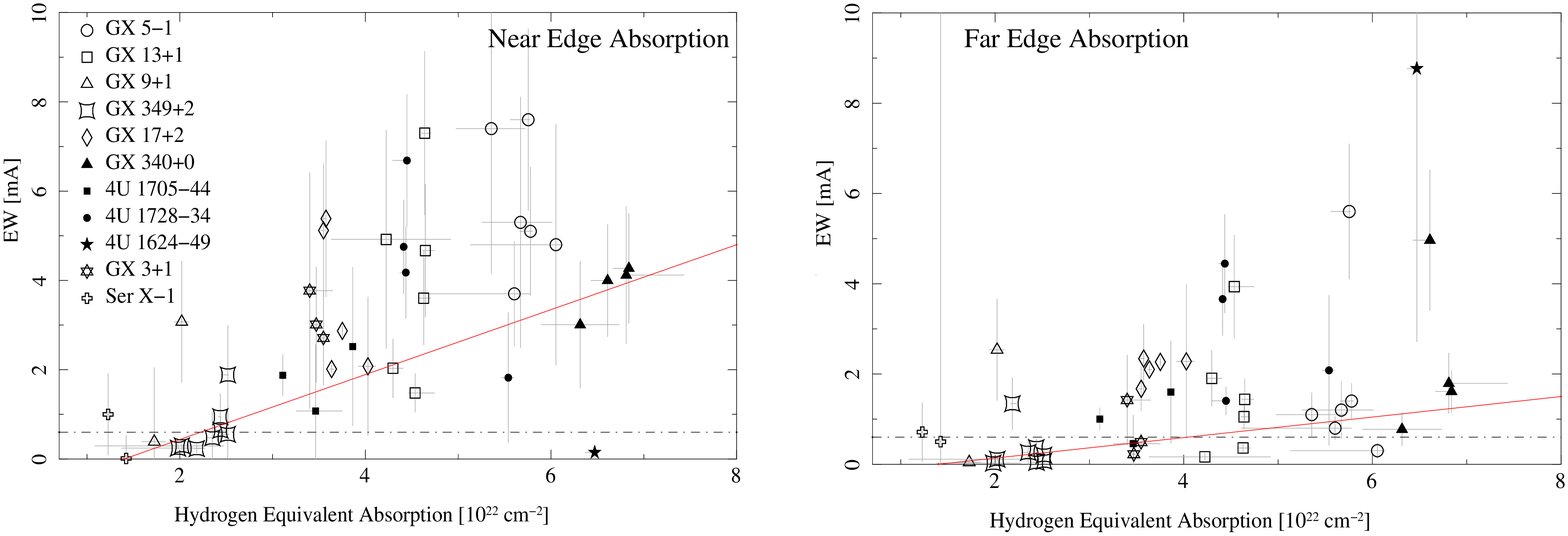} 
\figcaption{The equivalent widths of the near (left) and far (right) edge absorption
features with respect to Hydrogen equivalent column densities. The red line marks
a regression using only the smallest detected values in each source. {\bf The hatch-dot line
marks the detection limit of an equivalent width}.  
\label{fig:edgepars1}}
\end{figure*}

Figure~\ref{fig:edgepars1} (left)  shows the measured near edge 
absorption equivalent widths with respect to
column density for all survey observations. The
equivalent widths cover a range of values between 0.6 and 8 mA. The horizontal
hatch-dot line at 
0.6 mA marks the non-detection limit for most observations and any value below
that line is an upper limit. The red line marks a regression to the lowest values
in all sources. It is very close to what equivalent widths would we expect
from the models in Fig.~\ref{fig:lia_sik}.
There is a correlation of equivalent widths with 
column density in that the equivalent widths become larger with
increasing column density. This is not surprising as this
reflects the general behavior of absorbing power on the curve of growth. However,
there are many widths that
are above the expectation from the model and there is a significant amount of scatter.

\subsection{Far Edge Absorption\label{sec:faredge}}  

We also modeled the far edge absorption feature using
a Gaussian function. The centroids
are plotted in Figure~\ref{fig:standard} as unfilled squares. The values center tightly 
around 1.865+/-0.001 keV (6.648$\pm$0.005~\AA). The straight line fit is 
the location of the rest wavelength of resonance line
of He-like \sixiii\ as listed in \emph{APED}\footnote{http://www.atomdb.org} 
and \emph{XSTAR}\footnote{http://heasarc.nasa.gov/docs/software/xtar/xstar.html}. 
The actual fit to the values is almost identical to this line and we identify this
feature with \sixiii.  

The measured equivalent widths (see Figure~\ref{fig:edgepars1} right)  
range between 0.6 and 8 eV and are similar to the amounts observed 
in the near edge absorption feature. In the
case of the far edge feature, however, many values are not significantly detected and 
measurements scatter near or below the detection threshold. The 
correlation of equivalent width with column density is much less pronounced
compared with the near absorption case, which is expected most of that feature
comes from plasma material close to the sources.

\subsection{Variability and Source Differences\label{sec:variability}}

We noticed that even though in many sources the general appearance of the Si K edge 
seemed quite similar, there were also some major differences. Most apparent was the difference
between 4U 1624-49, which hardly shows any structure, i.e. the case A optical depth 
is equal to the case B optical depth, and GX 5-1, which perhaps exhibits
the most prominent pattern, i.e. the case A optical depth is significantly smaller
than the case B optical depth, even though both sources have large columns.

When we compare different observations of the same source, we find that both, near and far
edge absorption appear variable. This is the case in at least half the sources in the sample.
This variability cannot be explained by
differences in observing and detector modes; in many cases the variability in the pattern is present 
while a source is observed at similar source flux and 
in the same detector mode. We searched for correlations with observed source flux
and found none. More important here is 
the slope of the continuum well above the Fe K edge, because these photons are 
mostly responsible for Si K shell ionization. However, we do not have statistical significance
in the HETG data in this specific bandpass. Unfortunately, we also cannot use the 
0th order spectrum because it is so heavily piled up,
and therefore we cannot investigate this issue further with our available data. 
In the following we focus on some prominent examples.

\begin{figure*}
\includegraphics[angle=0,width=16cm]{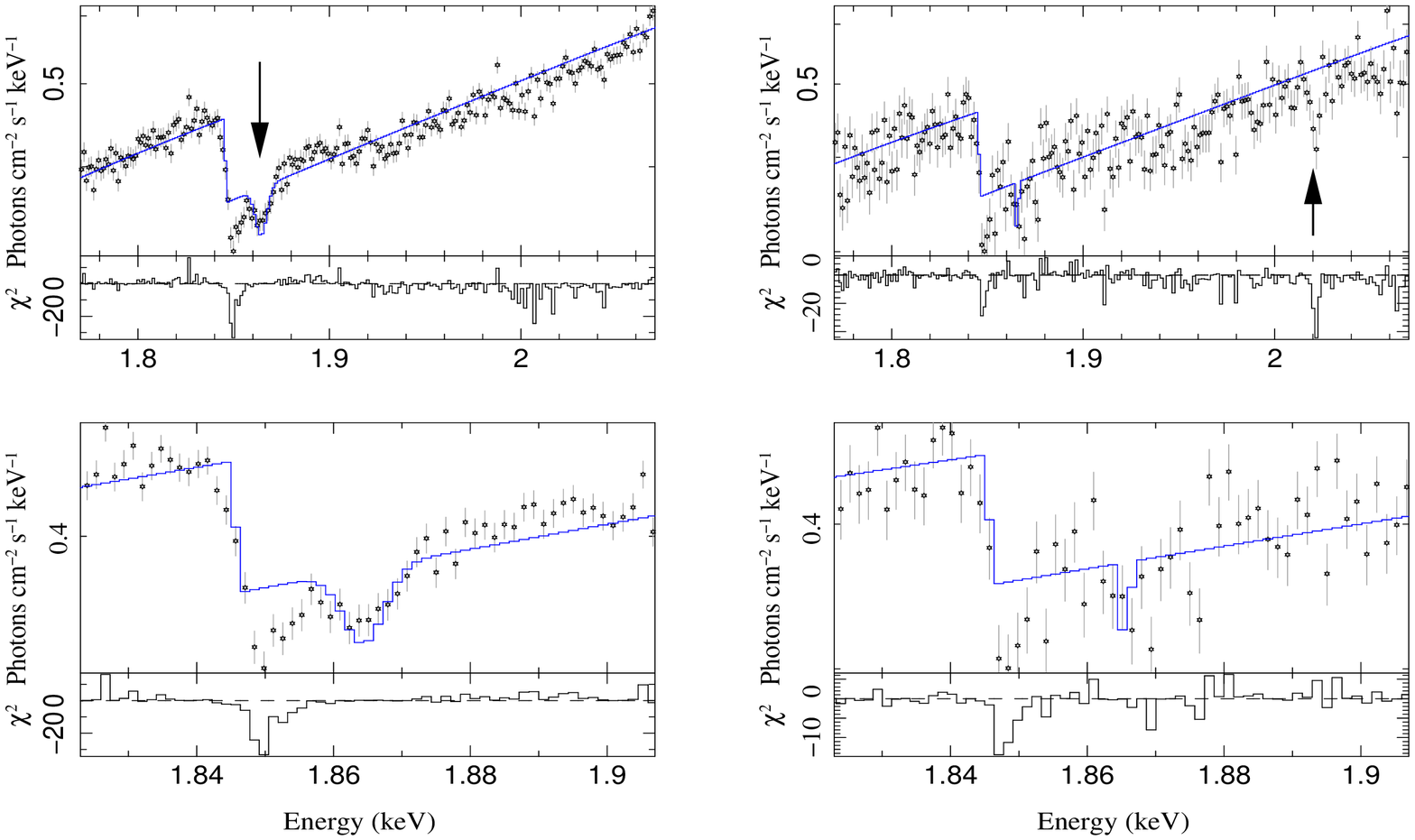}
\figcaption{The Si K edge region in GX 5-1 on a keV scale. In the diagrams on the left
(observation id 5888) the edge was fitted using an additional Gaussian function
at the location of the \sixiii\ resonance line at 1.865 keV line showing
residuals of the near edge absorption at 1.849 keV as well as some spurious
residuals at the \sixiv\ resonance line at 2.008 keV. In the diagrams on the right
(observation ids 10691-4) the edge was fitted with the same model.
The difference now is the there is barely a contribution at 1.865 keV,
but a strong one at the \sixiv\ resonance line location.
\label{fig:gx51sik}}
\end{figure*}

GX 5-1 is the brightest source in the sample with flux values exceeding 1 Crab. There are 6 observations
in the \chandra archive. The first observation (Obsid 715) appears to fit quite well
to the established silicate dust models as was shown by ~\citet{ueda2005}.
Other observations of this source are different, though.
The panels on the left in Figure~\ref{fig:gx51sik} show probably the best example
of a multi-peaked K edge structure as observed in Obsid 5888. The fit 
shows the edge and the far edge absorption
feature leaving the near edge absorption feature as a residual. The right panel shows 
a sum of observations a few years later, i.e. Obsids 10691-4. While the former 
features a strong far edge absorption feature, the latter in contrast show very weak far edge absorption.
Interestingly here we see significant \sixiv\ absorption. This 
indicates a shift in ionization parameter of atomic Si.
In Obsid 5888 we measure an equivalent
width of 5.6 m\AA\ for the \sixiii\ resonance line, 
while in Obsids 10691-4 the average equivalent width is very close to 
the detection limit of 0.6 m\AA. The near edge absorption, 
on the other hand is significant in all observations,
but the equivalent widths vary from 3.8 to 7.6 m\AA\ with a 90$\%$ uncertainty of about 0.5 m\AA.

\begin{figure*}
\vspace{0.2cm}
\begin{center}
\includegraphics[angle=0,width=16cm]{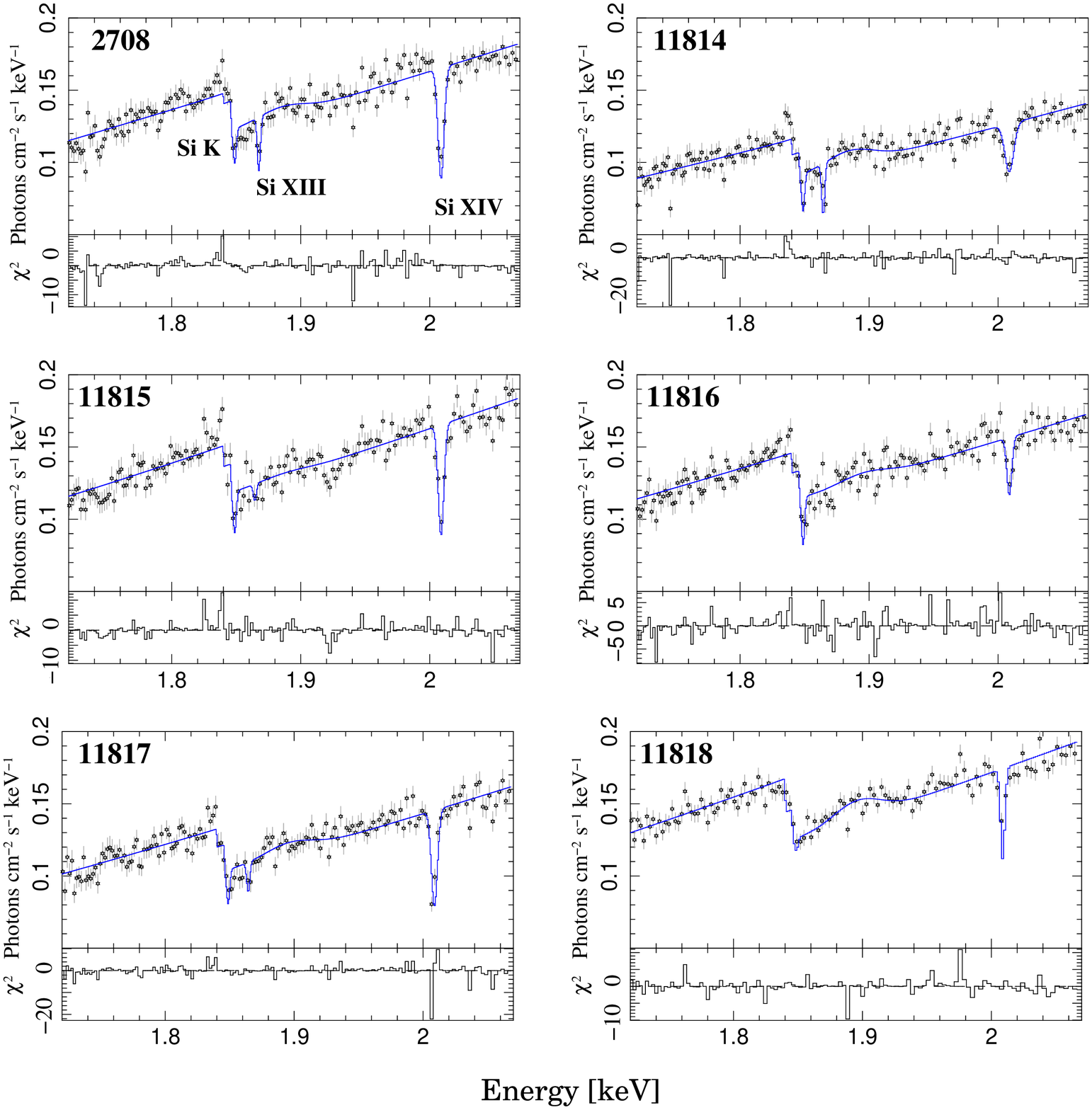}
\figcaption{The Si K edge in GX 13+1. The panels show the six HETG observations 
to date. All observations, except OBSID 11818, were taken using the same
timed event mode configuration. The fits here include all features described
in Sect.~5 below. The plot demonstrates the variable appearance of the
Si K edge specifically with respect to the line absorptions. The residuals at the edge
locations in the TE mode data likely stem from the scattering contribution to the 
optical depths (see Fig.~\ref{fig:lia_sik}). OBSID 11818 is in CC-mode which also
has lower spectral resolution and this feature is likely smeared. 
\label{fig:gx131sik}}
\end{center}
\vspace{0.2cm}
\end{figure*}

Perhaps the most intriguing source is GX 13+1 where we have six very similar observation
segments distributed over weeks and years. Figure~\ref{fig:gx131sik} shows these observations in sequence.
Obsid 2708 was the one used by ~\citet{ueda2005} in their Si K edge analysis. Readily detected is 
strong H-like \sixiv\ absorption stemming from an accretion disk wind reported by ~\citet{ueda2004} also
using Obsid 2708. 
However, we find that the Si K edge structure is highly variable
in the observations of GX 13+1 and while we can apply the dust model from ~\citet{ueda2005} in Obsid 2708 with
some minor residuals, we cannot reproduce such a fit in the other observations. 
We now identify the far edge absorption feature with He-like \sixiii\ resonance 
absorption and observe variations in equivalent widths
from $< 0.6$ m~\AA\ to about 4 m\AA. 
The 90$\%$ uncertainties are
of the order of 0.7 m\AA. Near edge absorption is present in all observations, however again 
with significant variations in equivalent widths. Equivalent 
widths vary from 1.4 m\AA\ to 7.4 m\AA.

For the two sources, GX 5-1 and GX 13+1, we also fitted a warm absorber model
to the Si K edge data where we fixed the \emph{TBnew(1).nh} parameter 
to the values in Tab.~1 . For the warm absorber model we use the \emph{XSTAR}
function \emph{warmabs} in \emph{XSPEC}.
The parameters of the near edge absorption function we fixed to the near edge locations
and to the observed case B optical depths with a width equivalent to the HEG resolution.
Such a function mimics the a basic Si K edge structure.
The ionization parameter $log \xi$
was constrained to values between 1.6 and 2.0, here the \sixiii\ line absorption is strong and
\sixiv\ is still weak. In GX 13+1 we added a warm absorber with $log \xi = 3$
in order to account for the apparent \sixiv\ absorption feature from a known disk
wind~\citet{ueda2004}.
We tested the warm absorber model for three specific cases, GX 5-1 (Obsid 5888),
GX 13+1 (Obsids 2708, 11814), GX 13+1 (Obsids 11815, 11816). In the last, \sixiii\ is
hardly detected and here we see the least ionized case and a more accurate silicate edge structure.
The warm absorber needs high turbulent velocities ($>$ 700 \kms) in order to fit the full \sixiii\ feature
width but also to smooth out small absorption features which are nested in the continuum noise.
This fact makes the fit very difficult because this creates many shallow minima in
the fit parameter space and allows for multiple solutions. In that respect such a simple
\emph{warmabs} fit is not unique. This effect was minimized by constraining the 
ionization parameter (log $\xi$) to 1.6 on the low, and 2.0 to the high side for finding the best fit.
To compute 90$\%$ confidence limits, we then released the constraint at got values for GX 5-1 of 1.8$\pm$0.3 and
for GX 13+1 of 1.9$\pm$0.3. If we would not
constrain the fit first, lower values down to 1.1 were found as well. In general,
these fits are able to provide
for the major features and their variability near the Si K edge.

Variations of EW span similar ranges among different sources and observations.
The source 4U 1624-49 
is different in a sense that it is the 
only one where we do not observe significant near edge absorption, but high far edge 
absorption which stems from an identified warm absorber in the accretion disk~\citep{xiang2009}.

Ser X-1 is the source with the lowest overall column ($1.2\times 10^{22}$ cm$^{-2}$). An early
\chandra observation was pileup dominated and mostly useless (OBSID 700), but we carried out
two observations in TE mode with a very narrow subarray (134 rows, OBSIDs 17485, 17600) 
reducing pileup to less than 3$\%$ at the Si K edge. Here we do not observe any edge structure 
in the MEG spectrum and only a very weak and unresolved near edge feature in the HEG. 

\section{High Definition HEG Modeling\label{sec:highres}}

There are a few observations in the sample that provide a specifically high brilliance 
view of the Si K edge region. MEG data do not provide enough resolution to
see possible super-fine structure in the edge. In first order the HEG provides a spectral
resolution of $E/{\Delta}E$ = 610 at the Si K edge, which might be enough to
detect some of this substructure. 
GX 3+1 at a moderately high flux and column density was observed in configurations
which kept pileup to less than 3$\%$ in TE mode. In addition to the fit model described 
in Sect.~\ref{sec:survey} we added some components local to the Si K edge
necessary to remove fit residuals. One
component is approximated by a broad Gaussion emission function to correct a slightly 
curved edge recovery centered at 1.88 keV (6.59~\AA) (see Fig.~\ref{fig:gx131sik}) and
which does not appear obvious in the extinction models in Fig.~\ref{fig:lia_sik}. Other
components involve a narrow Gaussian peak on top of the edge, as well as a second standard
edge function at slightly lower energies.

Figure ~\ref{fig:gx31edge} shows the Si K edge in the HEG summed first orders in the 
observations of GX 3+1. Several elements
are identified: a peak on top of the edge stemming from X-ray scattering
(see Fig.~\ref{fig:lia_sik}), a possible edge
at 1.840 keV, the main edge at 1.845 keV, and a weak \sixiii\ absorption line.
The scattering peak is seen in most of the edges, the possibility of more than
one edge is also hinted in some of the GX 13+1 edges (see Fig.~\ref{fig:gx131sik}).
The edge in GX 3+1 shows a two edge structure, with an optical depth ratio
below 0.1, an about 10$\%$ contribution to the measured optical depths above.  
The effect is currently barely at the 3 $\sigma$ level, but it demonstrates that these
edges have substantial complexity. 

\section{Silicate Extinction  Modeling\label{sec:extinct}}

The previous sections have shown that the Si K edge in high resolution spectra
is not only very complex, i.e. they show various structures due to silicate properties, 
as well as neutral and ionized atomic contributions, but there is scatter due to  
systematic uncertainties. This makes detailed data modeling extremely difficult.
However there are some signatures and trends in our measurements which allow us
to do some global model comparisons.  
In this section we attempt to characterize the observed silicate dust 
properties in the context of the models shown in Fig.~\ref{fig:lia_sik}.
The most common signatures in all the data stemming from silicate properties, besides
the edge location, are the case A and B optical depths. Their values as well
as their ratio carry significant information about dust properties. The models
shown in Fig.~\ref{fig:lia_sik} show that scattering is an important contributor
to the final optical depths oberved in the edges. Scattering specifically becomes a 
dominant factor when we focus on the underlying grain size distributions. 
In terms of scattering, small grain
sizes produce less, and large grain sizes produce a greater contribution to the extinction optical
depths. Figure~\ref{fig:lia_sik} (bottom) shows how large grain sizes modify
the edge function to a much  
larger extent than the small grains, which produce an edge shape that is similar
to the MRN distribution.   

\vspace{0.2cm}
\includegraphics[angle=0,width=8cm]{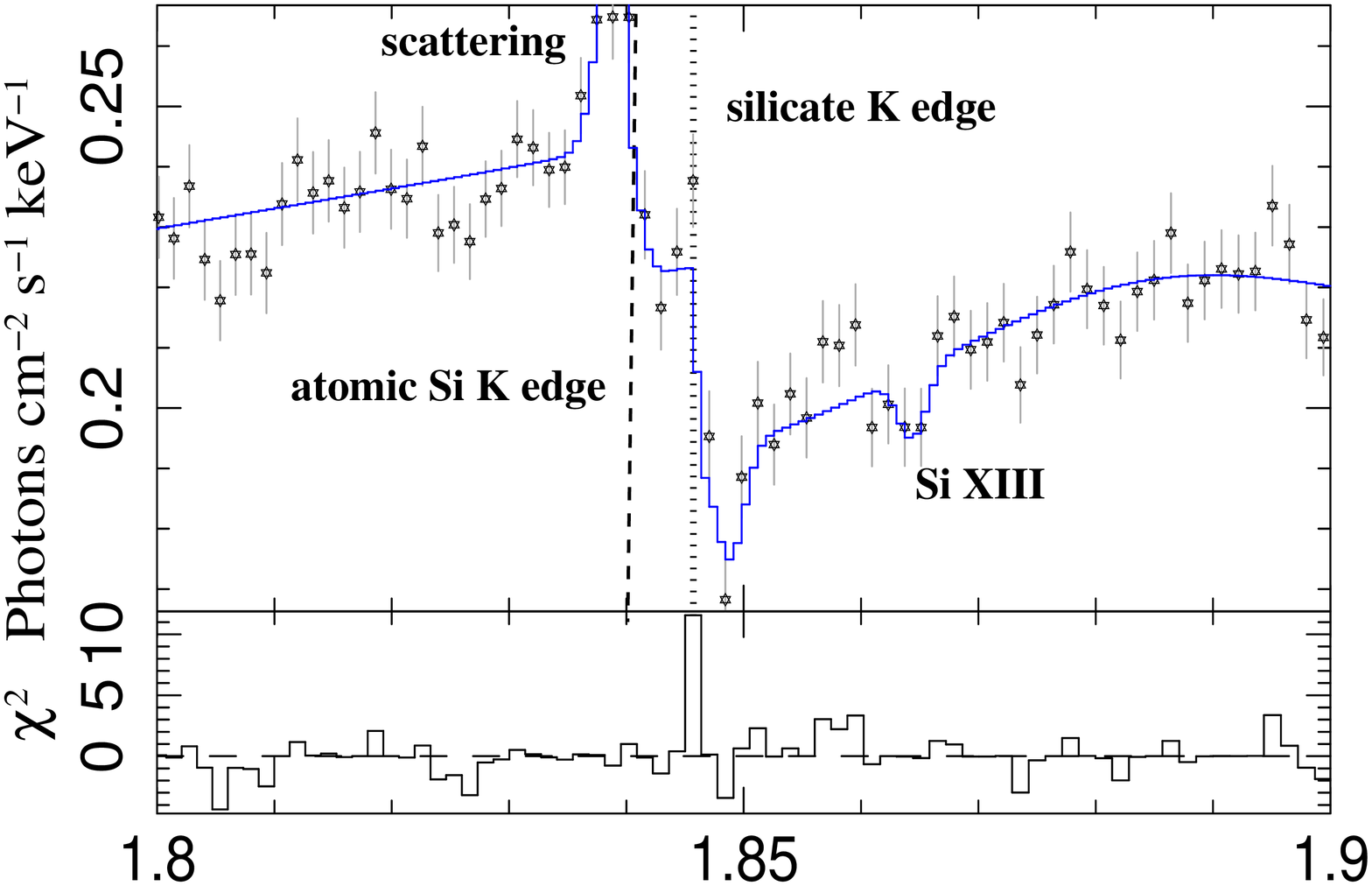}
\figcaption{The observed Si K edge in GX 3+1 summing up the HEG 1st orders
of Obsids 2745, 16307, 16492 with a total exposure of about 120 ks. Marked
are the contributions from scattering (see Fig.~\ref{fig:lia_sik}), a possible atomic Si K edge at 1.840 keV,
a silicate edge at 1.845 keV, and a weak absorption line from ionized \sixiii.
\label{fig:gx31edge}}
\vspace{0.2cm}

We created model grids for the two model examples described above, one
for the MRN grain size distribution which features predominantely small
($<< 0.1\mu$m) grain sizes (see Fig.~\ref{fig:modelgrids}, left diagrams)
and for large grain sizes of $0.3~\mu$m (see Fig.~\ref{fig:modelgrids}, right diagrams)
of Si dust.  
The black solid line refers to the model with the silicon abundance of ~\citet{wilms2000}
with all the Si in dust (i.e. a depletion factor of 1).
The green hatched lines represent the same as the black lines but 
with different relative overabundance factors.
Note that the model calculations include effects of self-shielding,
which reduces the edge optical depth for large grains.

When we compare the data to the model we see that in all cases the measured optical depths
generally follow the slope of the model tracks, which for case A optical depths is expected, 
however for case B optical depths it might not necessarily. 
It generally means that all lines of sight have fairly common
dust properties notwithstanding the data scatter. More importantly, however, as can be seen
for the MRN distribution model, the case A to case B optical depths do not
follow the same model tracks indicating that the observed ratio of the two depths is not 
consistent with an MRN distribution of dust or we do not have the correct model. 
The $0.3~\mu$m grain size model
provides smaller case B optical depths which is what we observe. Even though 
the case A to case B optical depths do not follow common model tracks in both 
shown cases, the two cases appear much more consistent when larger grain sizes are considered. 
A case for this can be made when we plot grain size effects into the right panel
of Fig.~\ref{fig:depth}. Here the observed case B optical depths appear
in between the expectation from the MRN distribution and the $0.3 ~\mu$m case. 
We likely observe significant contributions from scattering
of large grain sizes  and that future dust models have to consider that.   

\section{Discussion}

The Si K edge in high resolution X-ray spectra observed with \chandra shows 
very distinct structures. Next to the edge itself we observe broad near edge
absorption as well as broad far edge absorption. Both absorption features can
be fit with Gaussian line absorption. Due to the fact that there is so much additional
sub-structure, the placement of the edge location is difficult. This has to do with the
interpretation of the nature of the near absorption edge structure. In dust absorption
models and laboratory data of silicates such as presented in ~\citet{li1995} and
~\citet{costantini2014}, the near edge
drop in flux is very steep and we would place the onset of the 
edge itself right where it is observed. This value is shown in Figure~\ref{fig:standard}
at 1.844 keV. In most observed edges, however, there is
a near edge absorption feature which is consistent
with a Gaussian shape and cuts into the actual edge making
it appear slightly redshifted. Correcting for this effect would place the edge 
slightly higher near 1.845 keV. 

\begin{figure*}[t]
\includegraphics[angle=0,width=16.5cm]{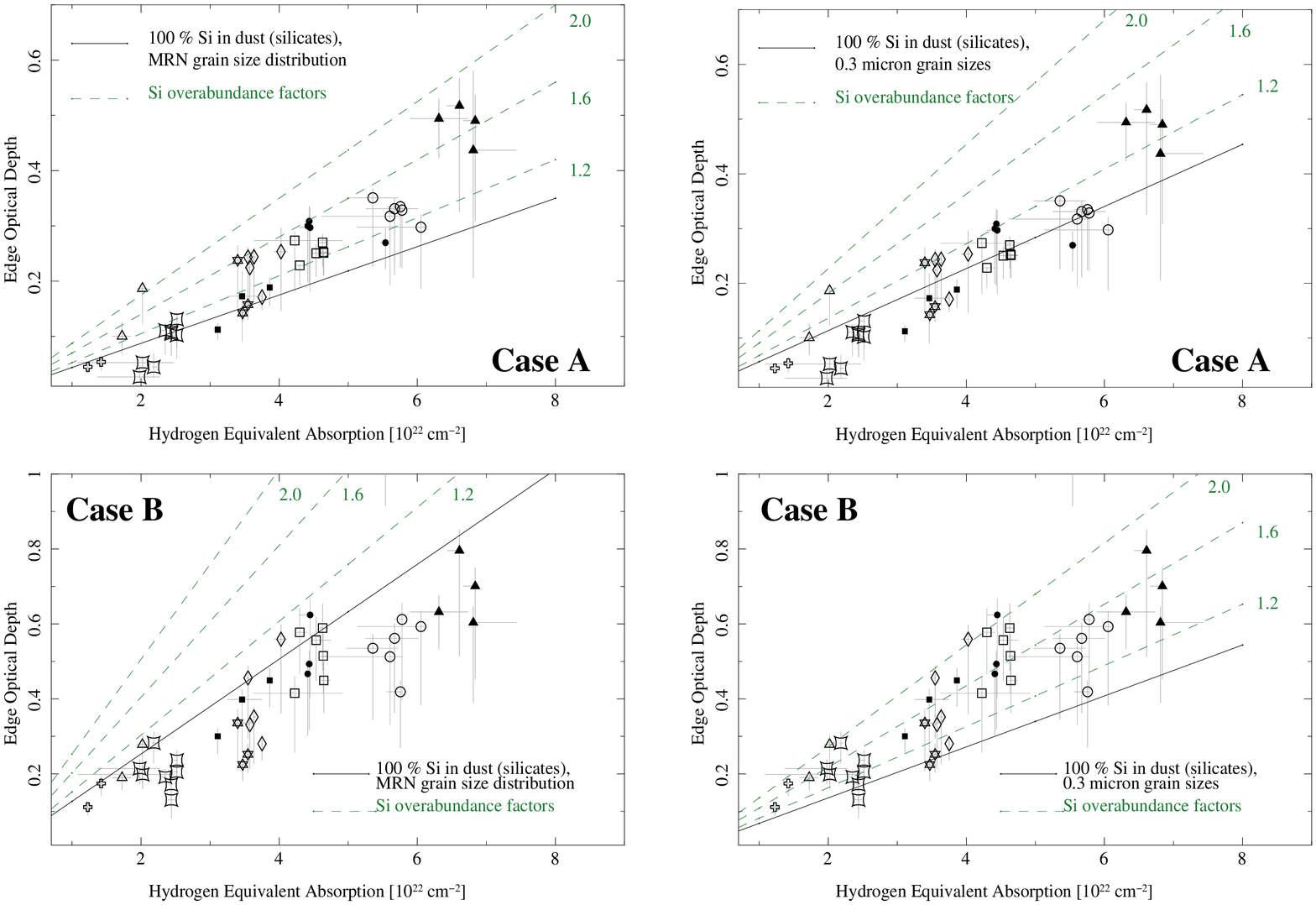}
\figcaption{The measured case A and case B optical depths and various extinction
models with respect to grain sizes, silicate fractions in dust as well as Si
overabundances with respect to the ISM Si abundance in ~\citet{wilms2000}.
{\bf Left:} Optical depth plotted over an extinction model grid calculated
for a MRN grain size distribution with 100$\%$ of Si in dust (solid line).
The green dashed line are the same plus a contribution to the optical
dephts from some Si overabundance with respect to the ~\citet{wilms2000} abundance.
{\bf Right:} The same approach as seen on the left but for 0.3 $\mu$m grain sizes.
\label{fig:modelgrids}}
\end{figure*}

Whether this edge location is truly a signature of silicates or is still consistent
with atomic silicon is a matter of debate. 
Literature values for atomic Si are some times significantly lower, 
~\citet{bearden1967}, for example,
has 1.838.9$\pm$0.4 keV, a value still commonly used to calibrate laboratory data
(see ~\citealt{drake2006}). 
In fact our measured values are lower but very close to the value 
of 1.8466 keV from \citet{verner1995} and 1.8486 from 
Hartree-Fock calculations~\citep{huang1976, gould1991} for atomic silicates. 
The former value has been adopted in the \emph{XSPEC} absorption
functions such as \emph{tbabs} and \emph{TBnew}, while the more commonly used \emph{wabs} and \emph{phabs}
still have the lower value implemented for atomic Si. However, we caution
the use of calculations for absolute wavelength determinations. In our previous studies
we always had to apply significant wavelength shifts to match calculations and \emph{Chandra}
measurements using benchmarked lines~\citep{juett2004, juett2006}. During the
ACIS instrument development phase in the mid-1990s, measurements of atomic Si in poly-silicon 
films based on synchrotron measurements of the \emph{Chandra} CCD gate 
structures~\citep{prigozhin1998} were performed to characterize the instrument
contribution of Si K absorption and here also the above value
of 1.839 keV was determined. This also marks the location of the instrumental Si edge 
in the uncorrected \emph{Chandra} data. The studies by ~\citet{ueda2005} and
~\cite{draine2003} rely on laboratory data from ~\citet{li1995}. More recently
newer laboratory X-ray studies of silicates basically confirm the expectation of
the Si K edge from silicates at the observed location of 1.844 keV by \emph{Chandra}~\citep{costantini2014}.  
If we adopt the hypothesis that the lower value is atomic and the higher value is from silicates, then
we believe that we see some atomic Si at the lower edge location and the bulk of Si K absorption
at the higher edge location from silicates.

The question still remains as to what exactly causes
most of the edge to emerge blue-shifted by about 5 - 6 eV with respect to the expectation from atomic Si. 
The consensus to date is that most observed Si is bound in silicates. 
Consequently the X-ray study by ~\citet{ueda2005} also has concluded that most of the edges origins
has to be silicates. However, our study shows that even though the edge location
is consistent with silicates, its observed structure appears more complex, and the models used by 
~\citet{draine2003} and ~\citet{ueda2005} do not provide a common fit to all the data. 

One of the first complications is that we observe
variablity with respect to the far edge feature which we identify as
\sixiii\ absorption local to the X-ray emitting binaries.
Absorption in the ~\citet{ueda2005} study was modeled as
a very shallow almost flat absorption trough as it would appear in some silicate \emph{XANES}.
In that study this was a consequence of limited and selective data choices aided by the fact that this absorption appeared
rather weak in these data sets. The examples provided here by many more
data sets from GX 13+1 and GX 5-1 show that this part is actually highly variable and 
more consistent with ionized silicon.  
The basic shape of the silicate edge has the near absorption feature superimposed on
a standard edge step shape, which is illustrated in the sequence shown in Fig.~\ref{fig:gx131sik}.
Here the edge in OBSID 11815 appears to have minimal structure which could allow to set limits 
on the non-variable structure of neutral silicates. 
All other structure features appear variable on relatively short time scales
and likely have their origins in ionization effects. 

In this respect the edge shape suggested in Fig.~\ref{fig:lia_sik} appears reasonable,
but needs some modification and context. The optical depths in both cases in Fig~\ref{fig:depth} 
appear systematically above the expectation from~\citet{wilms2000}. 
The mere fact that we observe more optical depth does not 
easily translate into a simple Si overabundance but depends on various
factors as how much Si is actually bound in dust, how much Si exists in
atomic and ionized form as well as details about dust properties such as
the choice of a dust size distribution. 
This fact 
has been recently emphasized by ~\citet{hoffman2015} and ~\citet{corrales2016}
by pointing out the importance
of scattering with respect to the Si extinction optical depth in the X-ray band.
Our preliminary attempt to compare our data with silicate models and various
dust distributions showed that we likely will not find a unique solution in the 
X-ray band. Fig.~\ref{fig:modelgrids} shows that even though larger grain 
sizes may be indicated,
there are additional unknowns such as the amount of Si in dust, 
the amount of atomic Si, a possible overabundance with respect to ISM abundances.
It is interesting to note, that even if we assume
that all the excess of Si in the \emph{case A} optical depths is due to an
overabundance, the amount is still slightly less that what we would expect from
solar abundances~\citep{asplund2009} for some of the data.

Our first attempt to observe more fine structure in the Si K edge has also
revealed that we observe another edge
at 1.840 keV, which is very close to the value expected for atomic silicon.
The effect is small, likely contributing less than 10$\%$ to the overall 
optical depth. This is the first time we have observed both, neutral atomic
silicon with the bulk of absorption from silicates. 
The fact that there is atomic silicon in the line of sight is also evident
from the presence of ionized atomic silicon. However the measured properties
of the ionized absorption suggests that it is of circum-binary nature. The identified
\sixiii\ resonance absorption shows velocities in excess of $\sim$ 700 \kms, which
cannot be part of the generic interstellar medium in the line of sight because
here we would expect velocities of much less than 100 \kms. Interestingly, we do not observe
resonance absorption from other elements such as \mgxi\ or \sxv, which might point
to a specific overabunance of Si even with respect to these other elements. 
When we model this component with a warm absorber model we determine moderate
ionization parameters again indicating a circum-binary origin.  
We can estimate the ionized column from the linear part of the curve of growth
which relates the equivalent width to the column density as

\begin{equation}
{W_{\lambda} \over {\lambda}} = 8.85\times10^{-13} N_z \lambda f_{ij},
\end{equation}

\noindent
where $N_z$ is the column of the ion species $z$, $f_{ij}$ the
oscillator strength, and $W_{\lambda}$ the equivalent width of the
absorption line (see ~\citet{schulz2002a} for details). 
For an oscillator strength of 0.747 ~\citep{nist2014} for the \sixiii\
resonance line the observed equivalent widths ranging from 0.6 mA to 5.8 mA 
then correspond to silicon columns ranging from $N_z = 2.1\times10^{15} \rm{cm^{-2}}$
to $N_z = 2.0\times10^{16} \rm{cm^{-2}}$. In terms of hydrogen equivalent columns using 
abundances from ~\citet{wilms2000}, these range from 
$N_{H} = 2.\times10^{20} \rm{cm^{-2}}$ to $N_z = 1.0\times10^{21} \rm{cm^{-2}}$ and
are a very small fraction with respect to the overall column density 
consistent with a local origin for these columns. 
However, when we combine these columns with the result for GX 3+1 for the 
fraction of the atomic Si column, which in that case would be around 
$3.0\times10^{21} \rm{cm^{-2}}$, then we find an ionized fraction of about
7$\%$, which is similar to ionization fractions observed for lower ionizations
of O and Ne in the warm ISM
~\citet{juett2004, juett2006}.

Even though the two component structure of the K edge consisting of
a cold and an ionized component has been found 
for lower Z edges, i.e. oxygen K and 
neon K~\citep{juett2004, juett2006, gatuzz2014}, the case for Si K is 
different. In the low Z case we observe mostly cold edge signatures from
atomic oxygen and neon as well as ionization signatures from a wide
range of ionization stages. The features did not appear variable
and rather consistent with contributions from the warm and hot phases
of the interstellar medium. In these cases there is also hardly any measurable
dust component involved, even though ~\citet{pinto2013} estimates the presence
of about 10$\%$ as forms of dust. The Fe L edge is different again as here 
the entire edge is likely various forms of dust~\citet{juett2006, lee2009}.
The Si K edge structure appears different again, as here the main cold component
is dust in form of silicates, while the ionized component is circum-binary material.
The hot phase of the interstellar medium is not hot enough to produce
collisional ionized Si and the observed velocities are too high.  

There are yet not many sources available that allow us to compare 
hydrogen equivalent columns from silicate optical depths with observations
in other wavelengths. However, two sources in our sample are also
present in the low Z survey by ~\citet{juett2006}, which include Ne K and Fe L edges. 
Ser X-1 and GX 349+2 have
low enough columns to also exhibit Ne K and Fe L edges in the X-ray spectrum.
Here we find that the columns from our Si K edges are very consistent with 
the ones found at Ne K, but Fe L columns appear about 30$\%$ lower. However,
when we compare the Fe L column found in GX 9+9 from that study with the ones found in
the recent study by ~\citet{corrales2016} which uses the dust model used
in our study, then there is a similar discrepancy. More studies are necessary 
to confirm these findings, however these few cases already demonstrate the
necessity of proper modeling of scattering from dust in the ISM.

One of the more interesting tasks is to compare the measured Si optical
depths with results from other wavebands. \citet{predehl1995}, for example, established
a relationship of optical extinction A$_V$ with X-ray column density based
on measurements of the scattering halos observed with \ros. Our sample 
compares fairly well with theirs with a somewhat larger scatter around that
relationship that they observe.

The Si K (Case A) optical depths also do not indicate significant Si overabundances
and, as shown in previous paragraphs, the columns are consistent with other elements
in the X-ray band. However, we also know from the ionization signatures in our spectra
that there are source intrinsic contributions. We then compared our Si K Case A columns 
to recent H~I surveys as the Leiden/Argentine/Bonn Survey of Galactic H~I 
by \citet{kalberla2005} and more recently the 
Parks Galactic All-Sky Survey (GASS) III by \citet{kalberla2015}. 
The column densities from the GASS survey are expected to be much lower than what
we expect from the Si K values because only a fraction of all the hydrogen is expected to be 
in H~I form. Surveys such as the WHAM survey of
ionized hydrogen ~\citep{haffner2003} as well as molecular surveys ~\citep{dame2001} show
significant contributions of both, H~II and H$_2$, in the direction of the
Galactic Bulge (see also \citealt{gordon1976}).
Here we find that the Si K optical depths follow the column densities from the GASS survey
quite well with the H~I values systematically lower by about a factor 3, which is
not unreasonable.
We also expect some line of sight variations due to ionized and molecular hydrogen.
However, what this comparison shows is that the silicate columns we observe trace 
well with columns from H~I surveys and that the intrinsic source contributions of less than 10$\%$ 
we observe are reasonable. In order to make more quantitative comparisons, the availability 
of more comprehensive surveys specifically for molecular
hydrogen is clearly needed. 

\subsection{Conclusions}

\begin{itemize}
\item{The Si K edge as measured with \emph{Chandra} has significant substructure and 
  consists of an edge at 1.844$\pm$0.001 keV, additional near edge absorption at 
  1.850$\pm$0.002 keV, and far edge absorption at 1.865$\pm$0.001 keV.}
\item{The Si K edge structure cannot be fit by a common silicon dust model as
  was done in a more limited previous study by ~\citet{ueda2005}.}
\item{The Si K edge structure in X-ray binaries appears highly variable
  in time scales of days to weeks and should be explained by local ionization effects.}
\item{The far edge absorption feature is identified as \sixiii\ resonance absorption in the 
  close vicinity of the X-ray binaries' accretion disk.}
\item{Variations in far edge absorption feature include the appearance of 
  \sixiv resonance absorption in same cases and are an expression of changing ionization parameters
   in circum-binary material.}
\item{The bulk of the Si K edge systematically appears at a value where silicate absorption is
  expected, however the associated near edge absorption also appears variable possibly indicating
  local ionization effects in silicon compounds.}
\item{Comparisons with extinction models show that the Si K edge optical depth is determined 
  by contributions from absorption and significant amounts of scattering.}
\item{A preliminary comparison of Si K edge optical depths with models using
  differenct grain sizes suggest that larger grain sizes are favored.}
\item{Data from highly resolved edges with good statistics indicate the presence of 
  an additional edge at 1.840 keV with about 
  10$\%$ of the optical depth likely coming from atomic silicon.}
\item{The Si K edge in X-rays then has several contributions to consider: absorption from
  atomic silicon and from silicates attenuated by considerable contributions of scattering optical
  depths and variable ionized absorption featues.}
\end{itemize} 

Future studies should focus on the variability and multiplicity aspect of the
Si K edge structure and sub-structure as well as scattering optical depths. 
We are in the process of gathering more
data less affected by instrumental effects in order to further study the 
possibility of multiple edge components. At the end future observations with the
upcoming X-ray spectroscopy mission \astroh will become instrumental and most
fruitful for these studies.
 
\acknowledgments
We thank all the members of the \chandra\ team for their enormous efforts,
specifically D. P. Huenemoerder and M. Nowak for easing
data processing and fitting procedures. 

\bibliographystyle{jwapjbib}

\end{document}